# Towards Safe and Comfortable Vehicle Control Transitions: A Systematic Review of Takeover Time, Time Budget, and Takeover Outcomes

**Kexin Liang[1], Simeon C. Calvert[1], and J. W. C. van Lint[1]**

## Abstract

**Objective:** This review systematically investigates the determination of sufficient time budgets in conditionally automated driving, focusing on the interplay between takeover time, allocated time budgets, and outcomes.

**Background:** Conditionally automated driving requires human drivers to resume vehicle control within limited time budgets when system limits are reached. However, the significant variability in drivers' takeover time (the time needed to regain control) poses challenges in balancing time budgets to avoid being too short (compromising safety and comfort) or too long (reducing driver alertness). Prior work lacks systematic exploration of sufficient time budgets across scenarios and drivers.

**Method:** Following the Preferred Reporting Items for Systematic Reviews and Meta-Analyses (PRISMA) guidelines, 100 articles are selected for review. Review papers are analyzed to extract overarching insights on time budget design, while primary empirical studies are examined to complement and validate the findings. We systematically synthesize evidence along the takeover sequence, covering (i) drivers' required takeover time, (ii) the time budgets provided by automated driving systems, and (iii) the resulting takeover outcomes. Based on the synthesis, we further discuss how the reviewed evidence informs the design of adaptive time budgets.

**Results:** This review synthesizes evidence on the takeover sequence and shows that takeover time varies significantly across drivers and driving contexts. Fixed time budgets cannot accommodate such variability. Adaptive time budgets that adjust to driver takeover demands are therefore promising. We propose an adaptive framework in which the time budget is defined as the sum of a predicted takeover time and an additional takeover buffer. To support this framework, the review identifies a qualitative inverted-U relationship among takeover time, time budget, and takeover outcomes: outcomes improve when the time budget exceeds the predicted takeover time, but deteriorate when the budget becomes unnecessarily long.

[1]Transport and Planning Department, Faculty of Civil Engineering and Geosciences, Delft University of Technology, Delft, The Netherlands

**Corresponding Author:**
Kexin Liang, Transport and Planning Department, Faculty of Civil Engineering and Geosciences, Delft University of Technology, Stevinweg 1, 2628 CN, Delft, The Netherlands.
Email: K.Liang-4@tudelft.nl

**Conclusion:** Structuring the takeover sequence clarifies trade-offs in time budget design. Adaptive approaches that estimate takeover time and allocate an appropriate takeover buffer to achieve targeted outcomes show promise yet warrant further investigation.
**Application:** Insights can optimize human-vehicle interactions for safe and comfortable takeovers, thus promoting public trust and acceptance of conditionally automated driving.

## Introduction

In conditionally automated driving, one primary safety consideration is the transition of vehicle control (ToC) from automation to human drivers. This ToC is critical in situations where automation's capabilities are exceeded, requiring drivers to promptly resume control within constrained time budgets as a safety fallback to minimize potential risks (On-Road Automated Driving ORAD Committee, 2021). However, the complexity of human-vehicle interactions can make ToCs demanding and potentially hazardous, leading to substantial variability in drivers' takeover time—the period required by drivers to regain control (Skrickij et al., 2020). This variability poses a significant challenge for conditionally automated driving systems (CADS), as they must provide sufficient time budgets to accommodate diverse driver needs during takeovers (Li, Wang, et al., 2021). Knowing what a sufficient amount of time is to perform ToCs is vital for driving safety and user comfort and is the focus of this research.

Review articles have emerged to provide insights on facilitating smooth ToCs, such as driving states during ToC (Lu et al., 2016), user interfaces (Kim et al., 2021), and takeover request designs (Miller et al., 2024). These reviews highlight the importance of designing CADS that align with drivers' cognition and demands. Zhang et al. (2019) performed a meta-analysis on takeover time and observed that longer time budgets that are available for drivers to resume vehicle control generally lead to longer takeover time and improved takeover performance when compared to shorter time budgets. This observation indicates that time budgets have significant influences on drivers' takeover time and takeover outcomes, which is also reported by McDonald et al. (2019) and Weaver and DeLucia (2022). Specifically, on one hand, tight time budgets can lead to short takeover times and poor takeover performance (Gold et al., 2018), as drivers (especially those who are sensitive to risk) tend to conduct overreactions. This can harm drivers' takeover experience and put them in serious danger of accidents (Gold et al., 2013). On the other hand, longer time budgets give drivers more time to respond to takeover requests and help them to achieve more stable takeover performance and lower accident rates (Wan & Wu, 2018; Yin & Pan, 2020). But if time budgets are excessively long, the safety of takeovers will not be improved significantly as time budgets increase (Huang & Pitts, 2022; Wan & Wu, 2018) because of the limitation of driver capabilities for fulfilling takeover tasks. Such excessively long time budgets can cause inefficiencies in takeovers and then in the general traffic system (Skrickij et al., 2020). Huang and Pitts (2022) stated that early requests for drivers to resume vehicle control can be taken as false alarms, which is dangerous if drivers decide not to respond. On these bases, we conclude that sufficient time budgets are vital for the safety and comfort of vehicle control transitions.

Therefore, a natural question to ask is:

> *How can time budgets be determined to sufficiently accommodate diverse scenarios and driver needs, ensuring safe and comfortable vehicle control transitions?*

To answer this open question, we conduct a systematic review that focuses on time budget, but also includes takeover time and takeover outcomes. While existing studies have investigated the relationships between time budgets and both takeover time and outcomes, there remains a critical gap in research addressing how to determine optimal time budgets that can sufficiently accommodate the diverse takeover needs across varying scenarios, particularly in the form of review papers. Additionally, the takeover process is an integrated sequence where the time budget is closely interconnected with takeover time (i.e., the lower limit of sufficient time budgets) and takeover outcomes (i.e., the consequence of the supplied time budgets for given takeover time). Studies examining related takeover time, takeover outcomes, and various factors impacting these elements, indirectly contribute valuable insights toward determining sufficient time budgets across diverse scenarios and drivers. However, a systematic review that examines these three elements as an integrated sequence is still missing. To fill in this gap, we break down our investigation on the determination of sufficient time budgets into three sub-questions:

> **Takeover Time:** *how long do drivers take to fulfill takeover tasks?*
> **Time Budget:** *how long do conditionally automated driving systems offer drivers for takeover tasks?*
> **Takeover Outcomes:** *do the offered time budgets suffice?*

Here, we use "takeover outcomes" rather than "takeover performance" to reflect a broader and more inclusive evaluation scope, considering the hierarchical relationship among physiological signals, psychological states, and takeover performance (Wang, Wang, et al., 2025).

This review aims to provide valuable insights to readers seeking a systematic overview of human-vehicle interactions during control transitions and requiring guidance on designing human-centered conditionally automated driving systems.

## Methodology

This systematic review exploits Scopus and Web of Science as they have relatively strict document inclusion criteria (Martín-Martín et al., 2019). Within these two databases, the search query is defined as TITLE-ABS-KEY ((human OR driver) AND (interact* OR cooperat*) AND ((auto* OR self) AND (driving OR vehicle OR car)) AND ((takeover OR transit*) control) AND (((takeover OR react* OR respon* OR lead) AND (time OR performance)) OR "time-budget")). The time scope is limited to 2010-2025, considering that from around 2010 onwards the number of studies on control transitions has steeply increased (Lu et al., 2016). As for document types, conference reviews are excluded because they add limited value when related conference papers are included. Additionally, only English articles are considered. This search strategy leads to 185 records in Scopus and 92 records in Web of Science. On this basis, a stratified process of filtering articles is performed following the guidelines of Preferred Reporting Items for Systematic Reviews and Meta-Analyses (PRISMA) (Moher et al., 2009). After removing 74 duplicates, an additional 103 records are excluded for two reasons: not focusing on the Human-Vehicle Interactions (HVIs) during takeovers, for example, constructing a framework for training automated driving models; and not discussing takeover time, time budget, or takeover performance, for example, propose a taxonomy for takeovers. This filtering process results in 100 articles remaining.

For the analysis of these 100 articles, this review adopts a combined method of umbrella review and systematic review. Specifically, six review papers (Hu et al., 2024; Hungund & Pradhan, 2023; Martinez & Huang, 2022; McDonald et al., 2019; Weaver & DeLucia, 2022; Zhang et al., 2019) are examined to provide a high-level understanding of HVIs during takeovers via the umbrella review.

Other involved research papers are also analyzed to supplement the results of the umbrella review. Although meta-analyses provide precise estimates of isolated effect sizes, the primary aim of this work is to develop conceptual structures and relationships (such as taxonomies, hierarchical dependencies, and design-relevant pathways) rather than to quantify average effects. These goals are better served by qualitative synthesis and structured comparison. Moreover, several of the included review papers already report rigorous meta-analytic findings on takeover time (e.g., Zhang et al. (2019)) and takeover outcomes (e.g., Weaver and DeLucia (2022)), making re-estimation of established effect sizes of limited additional value. With this combined analysis method, we aim to provide a comprehensive overview of the state-of-the-art research on takeover time, time budget, and takeover performance, which can facilitate the explorations of sufficient time budgets for safe and comfortable takeovers.

## Takeover Time

Takeover time refers to the amount of time that human drivers take to resume vehicle control from conditionally automated driving systems (Zhang et al., 2019). It is the minimum requirement to be met by sufficient time budgets (Marberger et al., 2018; Zhang et al., 2019); otherwise, drivers may not have enough time to complete the takeover task both safely and comfortably. In this section, we clarify the concept of takeover time adopted in this review in Section Concept of takeover time, examine the lengths of takeover time observed in empirical studies in Section Lengths of takeover time, and synthesize the determinants of takeover time in Section Determinants of takeover time.

### *Concept of Takeover Time*

Takeover process consists of multiple overlapping interval, such as takeover time, reaction time, and response time. These terms are generally used inconsistently in the literature (Huang et al., 2024; Zeeb et al., 2016). For example, the takeover time defined in Alambeigi and McDonald (2023) is called takeover reaction time in Wan and Wu (2018). To avoid potential misunderstandings of the temporal terminologies during takeovers, we examine the definitions of takeover time in the literature and clarify the specific concept adopted in this review. Studies employing alternative definitions are excluded from further analysis.

A well-accepted definition of takeover time is the interval from the initiation of a takeover request until the start of manual driving, according to ISO 21959:2020. "The start of manual driving" is generally interpreted in two ways. The first interpretation is the start of motor readiness to resume manual vehicle control, that is, the first behaviorally observable indication that the driver is physically re-engaging with the driving controls. The specific observable action varies by context and study protocol, such as pressing a manual driving activation button (Seet et al., 2022), placing hands back on the steering wheel (Diederichs et al., 2022), etc. The second interpretation is the moment when the driver starts to consciously control the vehicle, that is, the start of meaningful longitudinal/lateral inputs (Gold et al., 2018). Here, we adopt the second interpretation for the following two reasons. On the one hand, the first interpretation focuses more on drivers' physical behaviors, while Zeeb et al. (2016) and Yoon and Ji (2019) noticed that the time drivers take to achieve motor readiness does not change significantly in different takeover scenarios. This may be because motor readiness is more related to drivers' reflexive reaction (Zeeb et al., 2016), especially in time-critical scenarios (Markkula et al., 2016). Such an interval is beyond our research scope; On the contrary, the second interpretation focuses more on drivers' conscious behaviors which are the underlying reasons for how drivers respond physically (Merat et al., 2019; Xing et al., 2021). This second interpretation can help researchers to conduct in-depth investigations into the underlying mechanism of HVIs during takeovers and provide valuable insights into how to improve the safety and comfort of takeovers (Du et al., 2020a). Thus, this review defines the takeover time as the interval between the takeover request and drivers' first conscious input. Such first conscious input is generally indicated by the angle of the steering wheel exceeding 2 or the position of the braking pedal exceeding 10% (Gold et al., 2013; Li et al., 2019).

To illustrate the clarified takeover time, we describe a takeover process in a temporal sequence which is shown in Figure 1. The sequence begins with a conditionally automated driving system performing dynamic driving tasks (ISO 21959:2020) and maintaining control of the vehicle. During this phase, the human driver may engage in non-driving-related tasks, such as watching videos. When the system encounters a situation beyond its operational design domain (ISO 21959:2020), a "system boundary" occurs. The system then initiates a takeover request and performs safety fallback actions, such as deceleration, to manage the transition. The timing of this takeover request defines the available time budget before system disengagement. Upon receiving the request, the driver disengages from non-driving-related tasks and transitions to the cognitive driving loop to assume conscious control of the vehicle and start meaningful human control (Calvert et al., 2024). The duration of this transition is referred to as the "takeover time" and is the focus of this review.

## *Lengths of Takeover Time*

To address the first research question, that is, how long drivers take to complete takeover tasks, we review studies that report the length of takeover time in various scenarios. To our knowledge, there is no consensus in the literature about a general takeover time that can be applied to all situations (Xing et al., 2021; Zeeb et al., 2015). For example, Favarò et al. (2019) reported 1143 takeover cases from four vehicle manufacturers and noted that drivers' mean takeover time varied across manufacturers, ranging from 0.83 to 3.10 seconds. Based on the data reported in Favarò et al. (2019), we calculated an average mean takeover time of 1.14 s. Meanwhile, Zhang et al. (2019) analyzed 520 takeover cases from the literature. Results show that the mean takeover time across studies varies from 0.69 s to 19.79 s and the average mean takeover time is 2.72 s. To explore this further, we summarize the takeover time across 28 cases from three studies and list these examples in Table 1. This table details how takeover time varies across different driver characteristics and scenarios, capturing variations based on factors such as driver age, non-driving-related tasks (NDRT), urgency of the takeover scenario, traffic density, and takeover request modality (TORM).

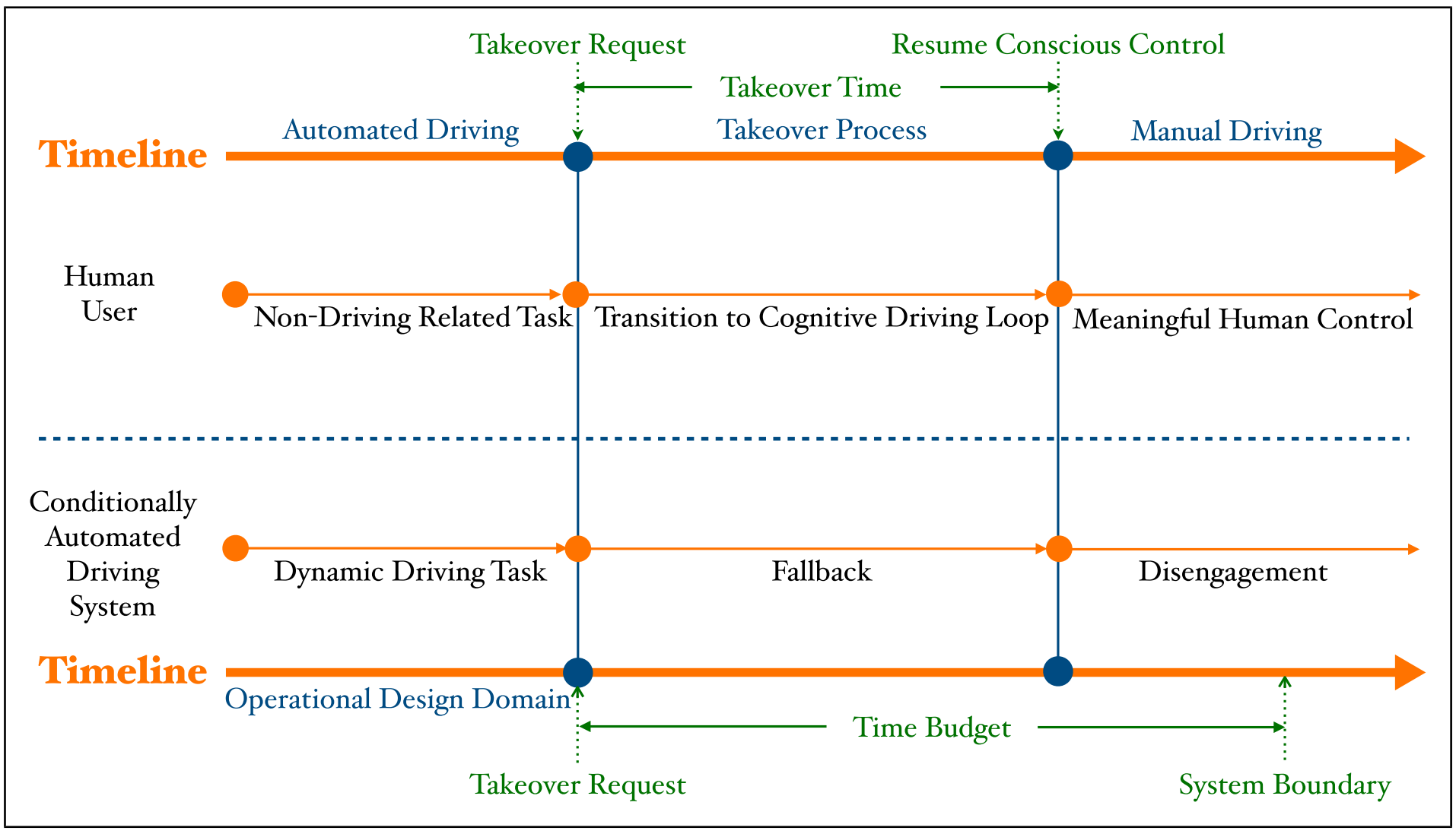

**Figure 1.** The timeline of a takeover process (from conditionally automated driving to manual driving)

Table 1 reveals significant variability in ToT across conditions. For instance, high-urgency scenarios generally prompt faster driver responses, while NDRTs (especially cognitively demanding tasks) tend to lengthen takeover time. There are also age-related differences: Li et al. (2018) found that older drivers typically require longer ToT than younger drivers under the same condition,

**Table 1.** An overview of takeover times studied in the literature (examples)

| Study | Age | NDRT | Urgency | Traffic density | TORM | ToT(s) |
|---|---|---|---|---|---|---|
| Körber et al. (2016) | Young | No | High | Low | A | 2.58 (0.97) |
| | Young | No | High | Medium | A | 3.32 (1.44) |
| | Young | No | High | High | A | 3.52 (1.17) |
| | Young | Phone conversation | High | Low | A | 2.76 (0.88) |
| | Young | Phone conversation | High | Medium | A | 3.70 (0.97) |
| | Young | Phone conversation | High | High | A | 3.66 (1.24) |
| | Old | No | High | Low | A | 2.41 (1.00) |
| | Old | No | High | Medium | A | 3.41 (1.34) |
| | Old | No | High | High | A | 3.41 (1.39) |
| | Old | Phone conversation | High | Low | A | 2.62 (1.29) |
| | Old | Phone conversation | High | Medium | A | 3.25 (1.41) |
| | Old | Phone conversation | High | High | A | 3.56 (1.10) |
| Li et al. (2018) | Young | Reading | Low | Low | V + A | 3.61 (1.79) |
| | Old | Reading | Low | Low | V + A | 4.33 (1.84) |
| Yoon et al. (2019) | All | Phone conversation | High | Low | V | 2.56 (1.10) |
| | All | Watching video | High | Low | V | 2.42 (1.09) |
| | All | Phone conversation | High | Low | T | 2.49 (1.35) |
| | All | Watching video | High | Low | T | 2.18 (1.33) |
| | All | Phone conversation | High | Low | A | 2.32 (1.29) |
| | All | Watching video | High | Low | A | 2.23 (1.27) |
| | All | Phone conversation | High | Low | V + T | 2.15 (0.91) |
| | All | Watching video | High | Low | V + T | 1.88 (0.80) |
| | All | Phone conversation | High | Low | T + A | 2.18 (0.99) |
| | All | Watching video | High | Low | T + A | 2.05 (0.99) |
| | All | Phone conversation | High | Low | V + A | 2.11 (1.08) |
| | All | Watching video | High | Low | V + A | 1.97 (0.95) |
| | All | Phone conversation | High | Low | V + T + A | 2.22 (1.07) |
| | All | Watching video | High | Low | V + T + A | 1.95 (1.07) |

*NDRT: non-driving related task; TORM: takeover request modality; A: auditory; V: visual; T: vibrotactile; ToT: takeover time, presented as "mean (SD)".

*Urgency of takeover scenarios is classified based on Zhang et al. (2019), where (i) low urgency: no foreseeable collision risk or a high time budget to collision (≥15 s); (ii) medium urgency: potential collision risk or disturbance to other road users if no response was made (e.g., the ego car drifting to the adjacent lane containing traffic), or a medium time budget (between 8 s and 15 s); (iii) high urgency: immediate risk of collision (time budget ≤8 s) if no response was made, or the participants were instructed to react to a stimulus as quickly as possible.

*Traffic density (TD) is manipulated by the number of surrounding vehicles per kilometer per lane, following Körber et al. (2016). Specifically, (i) low TD: ≤10 vehicles/kilometer/lane; (ii) medium TD: 10-20 vehicles/kilometer/lane; (iii) high TD: ≥20 vehicles/km/lane.

whereas Körber et al. (2016) showed no significant age difference in ToT. These two examples universally indicate that takeover time varies strongly over human drivers, driving contexts, and takeover scenarios (Delmas et al., 2022; Li et al., 2019; Wang, Xu, et al., 2025). Particularly, the comparison between Favarò et al. (2019) and Zhang et al. (2019) shows that takeover times observed in real-car experiments can differ substantially from those measured in driving simulators. This discrepancy is notable given that driving simulators are widely used to study human factors in automated driving (Du et al., 2020c; Gold et al., 2018; Zhao et al., 2023). Such differences should be considered when applying experimentally derived conclusions in practice.

Mean takeover times and their ranges provide a general indication of response duration. However, they do not capture the substantial variability in takeover time consistently reported across studies (Delmas et al., 2022; Li et al., 2019). Empirical evidence shows that takeover time distributions are typically right-skewed (Eriksson & Stanton, 2017b; Rydström et al., 2022; Zhang et al., 2019). This implies that a non-negligible proportion of drivers require much more time than the average to regain control. Designing time budgets based only on mean values therefore risks underestimating the needs of slower responders. Such underestimation may place these drivers in unsafe situations, with insufficient time to perform evasive actions. These findings underscore the need to move beyond averages and to focus on the sources of takeover time variability. Understanding this variability is essential for designing time budgets that better accommodate heterogeneous driver needs.

## *Determinants of Takeover Time*

Previous research has shown substantial variation in takeover time across drivers, driving contexts, and scenarios (Delmas et al., 2022; Wang, Xu, et al., 2025; Zhang et al., 2019). This variability complicates the determination of sufficient time budgets, which must account for takeover time as a lower bound. A clearer understanding of the factors influencing takeover time is essential for designing appropriate time budgets and improving takeover performance.

A growing body of studies on takeover time determinants has been executed on topics such as situation awareness (Chen et al., 2024; Yang et al., 2020), workload (Eriksson & Stanton, 2017b; Oh et al., 2024), time budget (Li, Wang, et al., 2021), non-driving related task (Hu et al., 2024). We notice that even for those studies that examine the same determinant(s), their experimental settings vary and consequently, their results can differ. These phenomena show that the studies of takeover time determinants are scattered and often lack comparability, which is also noticed by Zhang et al. (2019). We argue that two factors may be responsible for hindering comparative studies. First, determinants of takeover time are numerous (Zhang et al., 2019). It is unrealistic to investigate all determinants together. Second, these determinants are closely connected. Their interrelationships lead to potential confounding factors which make designs of controlled experiments challenging (Choi et al., 2020; Pipkorn et al., 2022). Therefore, it is difficult to isolate individual determinants from the overall experimental setups and to make quantitative comparisons of their effects on takeover time across studies.

This review extracts takeover time determinants through a combined umbrella and systematic review approach to ensure comprehensive coverage and methodological rigor. Specifically, we conduct an umbrella review to extract determinants of takeover time from existing review papers (Hu et al., 2024; Martinez & Huang, 2022; McDonald et al., 2019; Weaver & DeLucia, 2022; Zhang et al., 2019). These reviews form the foundation for compiling a broad set of determinants. We then perform a systematic review of additional empirical studies to complement the determinant set from the umbrella review (Agrawal & Peeta, 2021; Chen et al. 2021, 2024; Dogan et al., 2017; Du et al., 2020b; Kim et al., 2020; Li et al., 2022; Lin et al., 2020; Oh et al., 2024;

Roche et al., 2019; Ruscio et al., 2017; Stimm et al., 2019; Wu et al., 2019). This combined approach enables a comprehensive overview of the determinants of takeover time identified in the literature. After standardizing terminology and consolidating subcategories, 23 distinct determinants are identified and illustrated in Figure 2. Such an overview lays the foundation for investigating how takeover time determinants interact and the mechanisms through which they operate.

Here, we distinguish three closely related concepts: (i) drowsiness: a specific state characterized by a strong inclination to sleep. It degrades sensory processing and slows motor responses, potentially undermining takeover capability; (ii) fatigue: a broader condition of mental or physical exhaustion arising from high cognitive workload or prolonged passive monitoring (e.g., monotony/boredom). A driver may be fatigued without being drowsy; however, both can reduce the cognitive resources needed to regain situational awareness; (iii) arousal: the general level of physiological and psychological activation of the central nervous system. Low arousal can co-occur with fatigue/drowsiness, while high arousal may temporarily counteract sleepiness but can also impair performance if excessive (e.g., stress-induced overreaction). Overall, fatigue often leads to a decline in arousal, which, if sustained, can transition into drowsiness.

We find that the takeover time determinants from the five reviews have both overlaps and differences, mainly due to their different focuses. For example, Zhang et al. (2019) synthesized quantitative studies of takeover time. McDonald et al. (2019) were interested in driver behavior models during takeovers. Weaver and DeLucia (2022) focused more on the designs of conditionally automated driving systems. The determinants from other literature are mainly related to human factors, such as situation awareness (Agrawal & Peeta, 2021). This pattern highlights

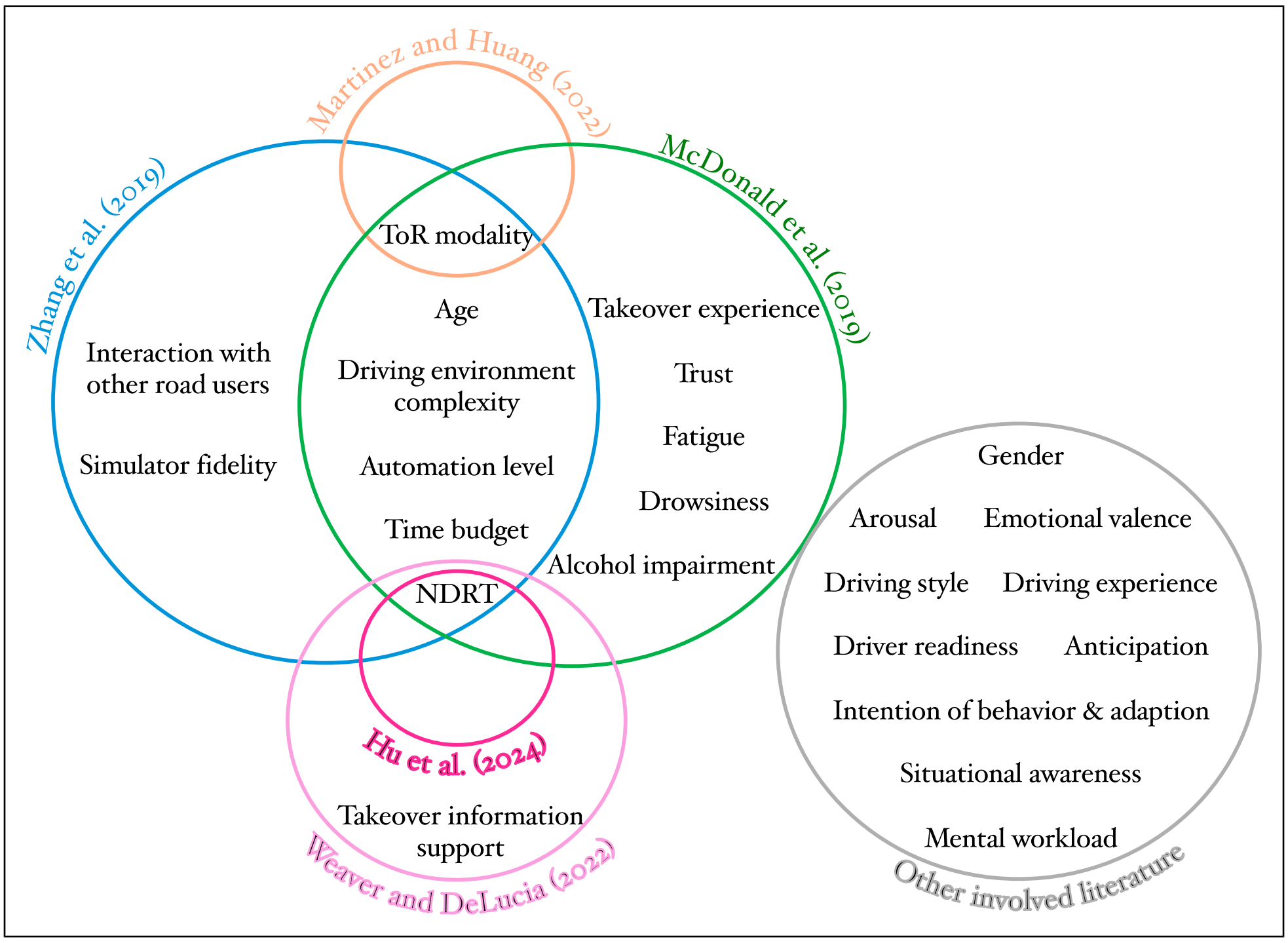

**Figure 2.** Overview of takeover time determinants

the need to better account for human factors when analyzing takeover time.

## Time Budget

Ensuring sufficient time budgets is critical for the safety, quality, and efficiency of takeovers (Li, Wang, et al., 2021), which are the fundamental aspects that influence drivers' willingness to utilize automation (Parasuraman & Riley, 1997). The sufficiency of time budgets is a relative concept, as it depends on the takeover time that drivers need to consciously control vehicle operations. On the one hand, the time budget should be longer than the takeover time. This is to ensure that drivers have enough time to respond and resume vehicles' control safely (Gold et al., 2018). On the other hand, if the time budget is excessively long, the efficiency and comfort of the takeovers can be affected, and drivers may perceive the takeover request as a false alarm (Huang & Pitts, 2022; Skrickij et al., 2020). This can also lead to potential risks to safe driving. This section delves into the sufficiency of time budgets by exploring fixed time budgets that are tested in the literature in Section Fixed time budgets and discussing the state-of-the-art adaptive time budgets research in Section Adaptive time budgets.

### *Fixed Time Budgets*

Existing studies on time budgets often focus on how fixed durations accommodate various takeover scenarios (Dogan & Acarman, 2025; Liu et al., 2024). A commonly used time budget in the literature is 7 seconds for distracted drivers to resume vehicle control (Gold et al., 2013; Jarosch et al., 2019; Körber et al., 2016; Xu et al., 2024). However, Gold et al. (2013) pointed out that drivers exhibit low situation awareness under a 7-second time budget, indicating its potential inadequacy. Similarly, Eriksson and Stanton (2017a) reported that a 7-second budget may not satisfy the needs of most drivers in real takeover contexts. These findings suggest that longer time budgets may be necessary, especially in complex scenarios or for drivers with slower response times. Further research is needed to identify context-dependent thresholds for fixed time budgets that ensure both safety and comfort during control transitions.

We dive deeper into the fixed time budgets that have been tested in empirical studies. Table 2 provides an overview of the suggested time budgets in different scenarios. The corresponding takeover time and driving velocities are also listed to provide contextual information. Such an overview can help readers to gain a better understanding of the relationship between time budgets and takeover time under various takeover contexts.

As shown in Table 2, the suggested time budgets in the literature vary significantly from 5 to 15 seconds. We describe three reasons responsible for such variations in suggested time budgets:

- **High variability of takeover demands**: empirical studies on fixed time budgets have diverse experiment designs, involving different participants and takeover scenarios. As discussed in Section Determinants of takeover time, this phenomenon leads to varied takeover times, that is, varied takeover demands that need to be satisfied by time budgets. So when researchers test the time budgets of interest, the sufficient time budgets that meet varied demands are accordingly different. This means that researchers can come to distinct conclusions about the suggested time budgets when they test different scenarios on different participants.
- **Lack of consistent measures of takeover outcomes**: different studies adopt different criteria for suggesting time budgets as they have different research objectives. For example, Yin and Pan (2020) reported that a time budget of 45s was preferred over 15s based on the subjective experiences and feelings of the elderly. Wan and Wu (2018) aimed to deliver a safe but also efficient takeover and they stated that a

**Table 2.** An overview of the time budgets studied in the literature (examples)

| Studies | tTB(s) | ToT(s) | V (km/h) | sTB(s) |
|---|---|---|---|---|
| Gold et al. (2013) | 5, 7 | mean = 2.10, 2.89 | 120 | ≥7 |
| Mok et al. (2015) | 2, 5, 8 | - | 72 | 5 |
| Melcher et al. (2015) | 10 | 1.4-6.7, median = 3.5 | 100 | 10 |
| Körber et al. (2016) | 7 | 1.91-4.28 | 120 | >7 |
| Clark and Feng (2017) | 4.5, 7.5 | mean = 2.07, 1.91 | 72 | 7.5 |
| Wan and Wu (2018) | 3, 6, 10, 15, 30, 60 | 2.0*-3.0* | - | 10 |
| Du et al. (2020a) | 4, 7 | 1.5*-3.5* | 130 | 7 |
| Dogan et al. (2021) | 4, 8 | mean = 2.28, 2.80 | 110 | 8 |
| Huang and Pitts (2022) | 4, 7 | mean = 1.75, 1.79 | 97 | 7 |
| Shahini et al. (2023) | 5, 8, 10 | mean = 2.05, 2.45, 2.60 | 64 | ≥8 |
| Wu et al. (2024) | 3, 7, 11, 15 | mean = 2.07, 2.98, 3.47, 3.08 | 60 | 15 |

*tTB: tested Time Budget; ToT: Takeover Time; V: Velocity; sTB: suggested Time Budget.
*The takeover times with * marks are estimated according to the figures in the references because the specific numbers are not mentioned.

10-second time budget is acceptable. This phenomenon reveals that standard measures for takeover outcomes are critical to improving the comparability of research on time budgets, which will be further discussed in Section Takeover outcomes.

**Sufficient time budgets should not be fixed**: time budget is the time to the detected system boundaries if the ego vehicle continues at its present speed on the same lane after the takeover request (Tanshi & Söffker, 2022). This reveals that the time budget is subject to the relative distance and driving speed to the detected system boundaries. Tanshi and Söffker (2022) found that a 7-second time budget is too long when the ego vehicle drives at 80 km/h, but is suitable when the driving speed is between 100 km/h and 130 km/h. This suggests that sufficient time budgets should not be applied uniformly, but adjusted based on the relative distance to and speed of approaching system boundaries.

In conclusion, fixed time budgets entail inherent limitations in satisfying the diverse demands of human drivers and takeover scenarios. A fixed time budget that is inadequate to cover the required takeover time may endanger the safety of takeovers, whereas an excessively long fixed time budget may impede the efficiency of takeovers. In either case, the takeover experience of drivers will be compromised. This can lead to reduced trust and acceptance of conditionally automated driving. Therefore, it is imperative to dynamically tailor the time budget based on the needed takeover time to improve the performance and user experience of takeovers.

## *Adaptive Time Budgets*

As fixed time budgets show weaknesses in satisfying various takeover demands (Eriksson & Stanton, 2017b), we argue that the adaptive time budget is a promising direction for research on time budget optimization. Adaptive time budgets can facilitate human drivers to perform flexible takeovers that are tailored to specific contexts. Previous studies also support this argumentation. De Winter et al. (2014) performed a meta-analysis of the effects of automated driving on drivers' workload and situation awareness. They stated that adaptive automation is a feasible approach to keep drivers in the driving loop. Huang and Pitts (2022) analyzed the effects of time budgets on takeover performance. The results suggested that the automated driving system should adjust the time budget based on the urgency of the situation. We argue that such adaptive time budgets have the potential to prompt better takeover outcomes and more human-centered conditionally automated driving compared with fixed time budgets.

Research on adaptive time budgets is emerging. The principal idea is to determine an adaptive time budget as a response time (typically takeover time) plus an additional buffer (Marberger et al., 2018; Li et al. 2021a, 2021b). In post-hoc analyses, takeover time is generally observed, and the extra buffer time is computed as the offered time budget minus the actual response time. However, from a time budget design perspective, takeover time must be estimated in advance, and the buffer time must be pre-defined before initiating takeover requests. These two aspects require further investigation to enhance the interpretability and practicality of the adaptive time budget strategies.

*Prediction of takeover time.* The studies of adaptive time budget have an implicit precondition: drivers' takeover time is known before determining the timing of initiating takeover requests. While some studies (e.g., Li et al.) did not mention that they predicted drivers' takeover time directly, the input variables they chose for the regression model are determinants and predictors of takeover time. Implementing adaptive time budgets requires accurate predictions of drivers' takeover time in real-time as a fundamental step. Therefore, this study synthesizes and compares different data streams and their validation status for estimating takeover time in Table 3.

In general, driver physical readiness and visual attention signals are consistently validated and deployment-ready predictors due to their direct linkage to takeover actions and low-latency acquisition. Context variables (operational complexity and automation features) improve robustness across scenarios and should be included when available. Psycho-physiological inputs are promising for

**Table 3.** Comparison of candidate input variables for takeover time prediction

| Category | Typical sensors | Validation status | Practical notes |
|---|---|---|---|
| Operational complexity | External vehicle perception sensors (e.g., Radar/LiDAR and GPS) | Widely validated | Improves Generalization across driving contexts |
| Automation features | System-level logs (e.g., time budget) | Widely validated | Critical indicators of system limitations |
| User profile | Questionnaires, historical records | Moderately validated | Static features; limited adaptability to momentary state changes |
| Driver physical readiness | In-cabin sensors (e.g., steering and pedal sensors) | Widely validated | Directly reflects motor readiness and action initiation capability |
| Driver visual attention | Driver monitoring systems (mainly cameras) | Emerging | Sensitive to occlusion and lighting; strong predictor of response delay |
| Driver cognitive state | Wearable physiological devices (e.g., heart rate tracker) | Emerging | Prone to environmental noise and motion artifacts |

*Note.* Validation status is defined based on evidence in the reviewed literature: (1) widely validated: characterized by robust, cross-study consensus where effects are consistently reproduced across diverse experimental setups; (2) moderately validated: supported by empirical evidence but characterized by context-dependent effects or mixed outcomes that suggest sensitivity to specific experimental conditions or limited replication; (3) emerging: represents areas of active exploration where preliminary evidence is growing, but standardized measurement protocols and a clear scientific consensus have yet to mature.

capturing latent cognitive states, but designers should treat them as optional enhancements and explicitly address signal reliability, calibration, and context sensitivity. To support transparent design choices, future work should report comparative evaluations (e.g., ablation studies and cross-context generalization tests) and compare deep models against interpretable baselines to quantify accuracy-interpretability trade-offs.

*Presetting the takeover buffer.* This review introduces the takeover buffer as a design-time parameter in time budget allocation. The takeover buffer is defined as the additional time intentionally added beyond the drivers' required takeover time, allocated before the vehicle reaches a system boundary. Therefore, the time budget can be expressed as:

$$\begin{aligned} \mathit{Time\ budget} = \mathit{Takeover\ time} \\ + \mathit{Takeover\ buffer}. \end{aligned} \tag{1}$$

Several temporal terms related to takeover buffer appear in prior literature, most notably time to the boundary at takeover timing (TTBT) (Li, Wang, et al., 2021) and maneuver response time (Tanshi & Söffker, 2022). These terms refer to the residual time remaining within the time budget after takeover time has elapsed:

$$\begin{aligned} &\mathit{Time\ to\ the\ boundary\ at\ takeover\ timing} \\ &= \mathit{Maneuver\ response\ time} \\ &= \mathit{Time\ budget} - \mathit{Takeover\ time}. \end{aligned} \tag{2}$$

Although these quantities seem to be numerically identical with takeover buffer, they differ fundamentally in conceptual role and usage. TTBT and maneuver response time are post-hoc, outcome-oriented measures used to assess whether a given time budget was sufficient once the takeover has occurred. In contrast, the takeover buffer is a proactive design variable specified a priori to determine how much additional margin should be allocated beyond the predicted takeover time. Introducing an explicit takeover buffer enables (i) tolerance to prediction errors in takeover time estimation, thereby improving the robustness of adaptive time budget strategies (Marberger et al., 2018), and (ii) mitigation of excessive time pressure on drivers during control transitions, supporting safer and more comfortable takeovers (Melcher et al., 2015). This design-oriented interpretation distinguishes the takeover buffer from existing residual-time constructs while remaining consistent with their mathematical definitions.

We leverage the records in Table 2 and subtract the mean/median takeover times from the suggested time budgets. Mok et al. (2015) is excluded because it did not report takeover time. The takeover time in Körber et al. (2016), Wan and Wu (2018), and Du et al. (2020a) is taken as the average of the maximum and minimum values because these two studies only reported ranges of takeover times. Results show that the mean takeover buffer is 6.00 s while the average mean takeover time is 2.65 s. This suggests that the optimal takeover buffer can be two to three times the takeover time, which accounts for a large part of the time budget and thus requires closer examination.

The optimal takeover buffer is generally determined via post-analyses after the takeover process based on the subsequent takeover outcomes Li, Hou, et al. (2021). For example, Li, Wang, et al. (2021) evaluated four takeover buffers (3, 4, 5, and 6 seconds), which they referred to as the "time to the boundary at takeover timing." The results indicated that a 4-second buffer was the minimum interval required to satisfy the preset collision rate. But from the perspective of adaptive time budget design, the takeover buffer should be preset and added to the predicted takeover time to determine the allocated time budget. This requires an established relationship between takeover time, takeover buffer, and the corresponding takeover outcomes (Gold et al., 2018; Tanshi & Söffker, 2022). Many determinants of takeover time are also likely to influence the required optimal buffer. When drivers require longer takeover times (often due to higher task demand or reduced driver capability), a larger takeover buffer is typically needed to support safe, high-quality control stabilization and maneuver execution. For example, under higher task demand (e.g., high traffic density), a driver with lower takeover capability (e.g., elevated cognitive workload) may require a larger buffer to achieve satisfactory takeover performance. Demographic factors (e.g., age and gender) and driver traits (e.g., driving style and trust in automation) may further shape buffer design by affecting driver comfort and trust in automated driving, which are key contributors to performance degradation when time budgets become unnecessarily long. These relationships remain underexplored and warrant systematic empirical validation.

## Takeover Outcomes

Takeover outcomes serve as key optimization targets in research on vehicle control transitions (Li et al., 2023), including studies on the determination of sufficient time budgets. This section synthesizes measures of takeover outcomes to clarify the suitability of various evaluation criteria. Specifically, we examine measures and the corresponding indicators of takeover outcomes in the literature in Section Measures of takeover outcomes, and compare the pros and cons of various human-related outcome measures in Section Comparison of human-related takeover outcome measures.

### *Measures of Takeover Outcomes*

Research on takeovers has extensively explored how various factors influence takeover outcomes (Hwang et al., 2025; Lin et al., 2020). These studies generally employ different evaluation indicators and thresholds (Li et al., 2023). For example, Eriksson and Stanton (2017a) employed the standard deviation of the steering wheel angle to indicate drivers' workload and used the mean absolute lateral position to assess drivers' lane-keeping accuracy. Alrefaie et al. (2019) quantified the quality of takeovers based on the mean percentage change of vehicles' speed and heading angle during the takeover process. These examples show that there are multiple measures of takeover outcomes, yet a well-accepted criterion is missing (Cao et al., 2021). This issue brings about a lack of comparability in research related to takeovers, which can introduce biases in quantitative studies of takeover outcomes and can even lead to opposite conclusions. To address this issue, it is essential to improve the clarity on the suitability of evaluation criteria for takeover outcomes.

A valid measure of takeover outcomes requires a systematic overview of performance indicators. This enables researchers to compare the pros and cons of these indicators and then develop standard mevaluation criteria (Li et al., 2023). To provide such an overview, we synthesize the studies of takeover outcomes and extract the related indicators via the combined approach of umbrella review and systematic review. Specifically, an initial indicator set of takeover outcomes is constructed by an umbrella review on four review papers, that is, McDonald et al. (2019), Cao et al. (2021), Weaver and DeLucia (2022), and Chen et al. (2025). Additional indicators from other involved literature (such as Gold et al. (2016), Beggiato et al. (2018), and Huang et al. (2024)) are also included to complement the indicator set. The applied indicator extraction process is shown in Figure 3, where 38 outcome indicators are identified. We acknowledge that several time-related indicators in Figure 3 (e.g., time to hands-on-wheel and time to feet-on-pedal) are also considered as takeover time in the literature. They can serve as event markers for operationally defining takeover time under specific protocols. In this review, we classify these indicators as outcome metrics, as they are used to evaluate takeover efficiency and quality in the referenced studies. Our classification follows this evaluative role, rather than being based solely on time units or their potential use as definitional endpoints.

These four review papers cover both vehicle-based metrics (e.g., lateral acceleration)

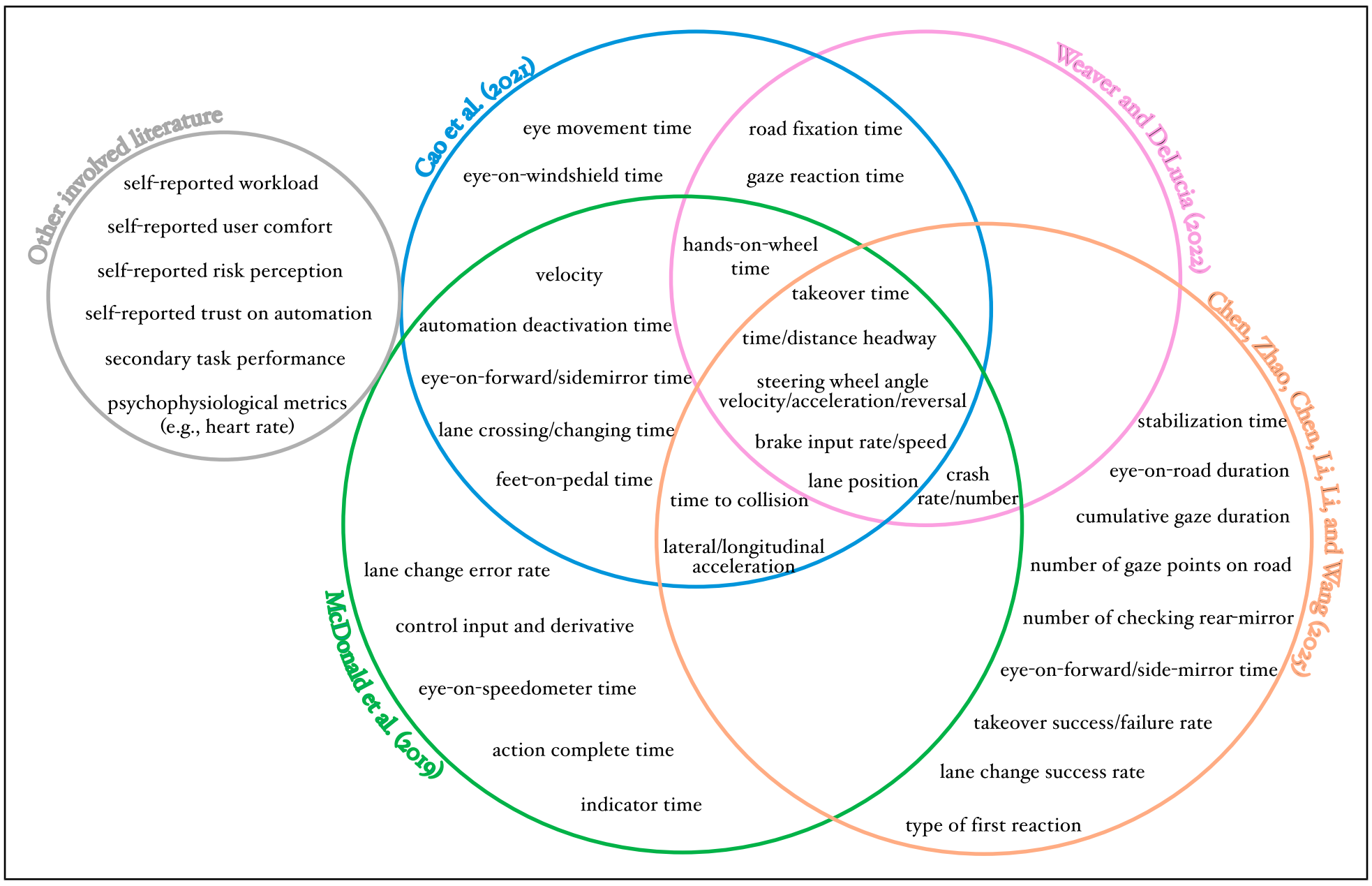

**Figure 3.** Extraction of takeover outcome indicators

and human-based reaction times (e.g., the time taken to return hands to the steering wheel). However, they place less emphasis on drivers' subjective experiences (such as workload and user comfort). Such indicators are primarily drawn from other involved literature to supplement the outcome indicator set. These experiential factors are crucial for driver acceptance of automated vehicles and ultimately influence users' willingness to adopt the technology. This phenomenon implies an imbalance in how objective and subjective outcome measures are being considered in the literature: some studies focus on driver experience (Li, Hou, et al., 2021, 2023), while others emphasize vehicle performance (Happee et al., 2017; Li et al., 2018; Lu et al., 2021). This imbalance leads to a fragmented view of what constitutes successful takeovers and can bias the related studies. For instance, Gold et al. (2016) considered a long minimum Time To Collision (TTC) as a marker of high-quality takeover quality. Minimum TTC is a valid metric from a vehicle-centric perspective, as it reflects a lower risk of collision. However, Radlmayr et al. (2018) argued that achieving a longer minimum TTC often involves abrupt braking, which can result in an unnatural and uncomfortable experience for the driver. Such discomfort may undermine trust in the automation and reduce long-term acceptance (Ma & Zhang, 2021). Using minimum TTC as a standalone indicator may be insufficient or even misleading from a human-centered perspective. Therefore, it is important to develop integrated takeover outcome measures that capture both the human perspective and the vehicle perspective to optimize the safety and comfort during control transitions.

## Comparison of Human-Related Takeover Outcome Measures

When it comes to takeover outcomes, the objects that are measured by human-related indicators are not as straightforward as those indicated by vehicle-related indicators. For example, drivers' heart rate and its variability are psycho-physiological responses that can be used to measure drowsiness (Fujiwara et al., 2018) and workload (Du et al., 2020a). Workload can also be evaluated via self-rating questionnaires like NASA-Task Load Index (Yoon & Ji, 2019). Taken the complex relationships between cognition constructs and human-related takeover outcome indicators, it is essential to clarify the suitability of these indicators and compare their pros and cons collectively. This will help to build a comprehensive understanding of human-related measures of takeover outcomes and identify suitable indicators for different research contexts.

**Table 4.** Comparison of human-related takeover outcome measures

| Category | Timeliness | Robustness | Implementability | Intrusiveness | Validity |
|---|---|---|---|---|---|
| Secondary task performance | **Medium:** processing depends on task completion; difficult for split-second real-time assessments. | **High:** reliable across scenarios if the secondary task is standardized. | **Medium:** easy to program, but requires the driver to be engaged in a specific task. | **High:** interferes with natural behavior by adding extra workload. | **Medium:** may alter the construct being measured. |
| Self-reported experience | **Low:** data collection is strictly post-hoc. | **Medium:** robust to withstand interferences as long as the questions are well designed. | **Low:** simple to implement via digital or paper surveys. | **Low:** typically does not distract drivers during takeovers. | **Medium:** captures subjective perception but subject to recall bias and self-assessment limits. |
| Psycho-physiological responses | **High:** enables real-time monitoring with millisecond-level data streams. | **Medium:** sensitive to noise and environmental conditions. | **High:** requires specialized sensors and signal processing. | **High:** often involves wearable or skin-attached sensors. | **High:** closely linked to mental states but requires careful interpretation and validation. |
| Behavioral indicators | **Medium:** processing depends on behavior completion; difficult for split-second realtime assessments. | **Medium:** affected by specific context and threshold definitions. | **Medium:** requires vehicle and driver monitoring sensors. | **Low:** generally drivers are not aware they are being measured. | **Medium:** objective but often surrogate indicators for driver cognition. |

In this section, we synthesize the analysis of various human-related measures of takeover outcomes based on the review paper of Hecht et al. together with other research articles such as Pakdamanian et al. (2021), Marberger et al. (2018), and Conti-Kufner (2017). Four categories of human-related measures (secondary task performance, self-reported experience, psycho-physiological responses, and behavioral indicators) are compared from five perspectives (timeliness, robustness, implementability, intrusiveness, and validity), as shown in Table 4.

Given that each measure has its pros and cons, we suggest that a hybrid approach that combines indicators from multiple categories can be valid for measuring the takeover outcomes from the human perspective. This approach is promising to reduce false detection rates by integrating multi-dimensional measurements. This is especially important when adaptive systems rely on potentially imperfect assessments of driver states. Our suggestion is in line with Hecht et al. (2018) and Lu et al. (2021). They concluded that a combination of measurements is necessary to capture the full range of driver state and driving performance, especially in complex scenarios.

Building on the above systematic comparison of human-related takeover outcome measures, we propose the following guidelines to support consistent and interpretable use of subjective outcome indicators in future research.

(1) Conduct cross-validation for subjective outcome measurements.

Given the complexity of human cognition, a hybrid approach that integrates indicators from multiple categories is recommended to enhance measurement interpretability and reduce false detection rates through cross-validations. This aligns with the conclusions of Hecht et al. (2018) and Lu et al. (2021), who emphasized that multimodal measurement is necessary to capture the full range of driver states, particularly in complex scenarios.

(2) Use validated and standardized instruments whenever possible.

The use of validated psychometric tools helps reduce measurement error, ensures construct validity, and greatly improves comparability across studies. Standardized instruments also facilitate meta-analytic synthesis and reduce the risk of construct drift, where different studies measure nominally similar constructs using inconsistent items.

(3) Advance validation of psycho-physiological measures.

Psycho-physiological responses exhibit particularly complex and indirect causal relationships with cognitive constructs and are highly sensitive to noise and contextual factors. Future research should (i) further investigate these relationships to provide clearer guidance for selecting appropriate physiological indicators and (ii) develop advanced signal-processing methods and measurement strategies to mitigate noise and contextual variability, thereby improving the robustness and generalizability of psychophysiology measures.

(4) Clarify hierarchical relationships among cognitive constructs.

Cognitive constructs such as workload, comfort, and trust are interdependent and should not be interpreted in isolation. Establishing and validating their hierarchical relationships through theory or latent-variable modeling (such as structure equation modeling) provides a consistent structural framework for determining which constructs must be measured, mapping indicators to the correct constructs, and interpreting subjective takeover outcomes in a standardized format.

## Discussion and Limitations

The sufficiency of time budgets is subject to the required takeover time and the corresponding takeover outcomes (Marberger et al., 2018). This systematic review has synthesized the existing research on takeover time (Section Takeover Time), time budget (Section Time Budget), and takeover outcomes (Section Takeover outcomes). We suggest adaptive time budgets as a potential solution to accommodate diverse takeover demands across various driving contexts. That is, conditionally automated driving systems predict drivers' takeover time before takeovers start, and then find the optimal time budgets according to the relationship between takeover time, time budget, and takeover outcomes. To facilitate the application of adaptive time budgets, we propose: (i) a taxonomy of takeover time determinants; (ii) a taxonomy of takeover outcome measures; (iii) a qualitative relationship among takeover time, time budget, and takeover outcomes; (iv) a research agenda for sufficient time budgets; and (v) a discussion of study limitations.

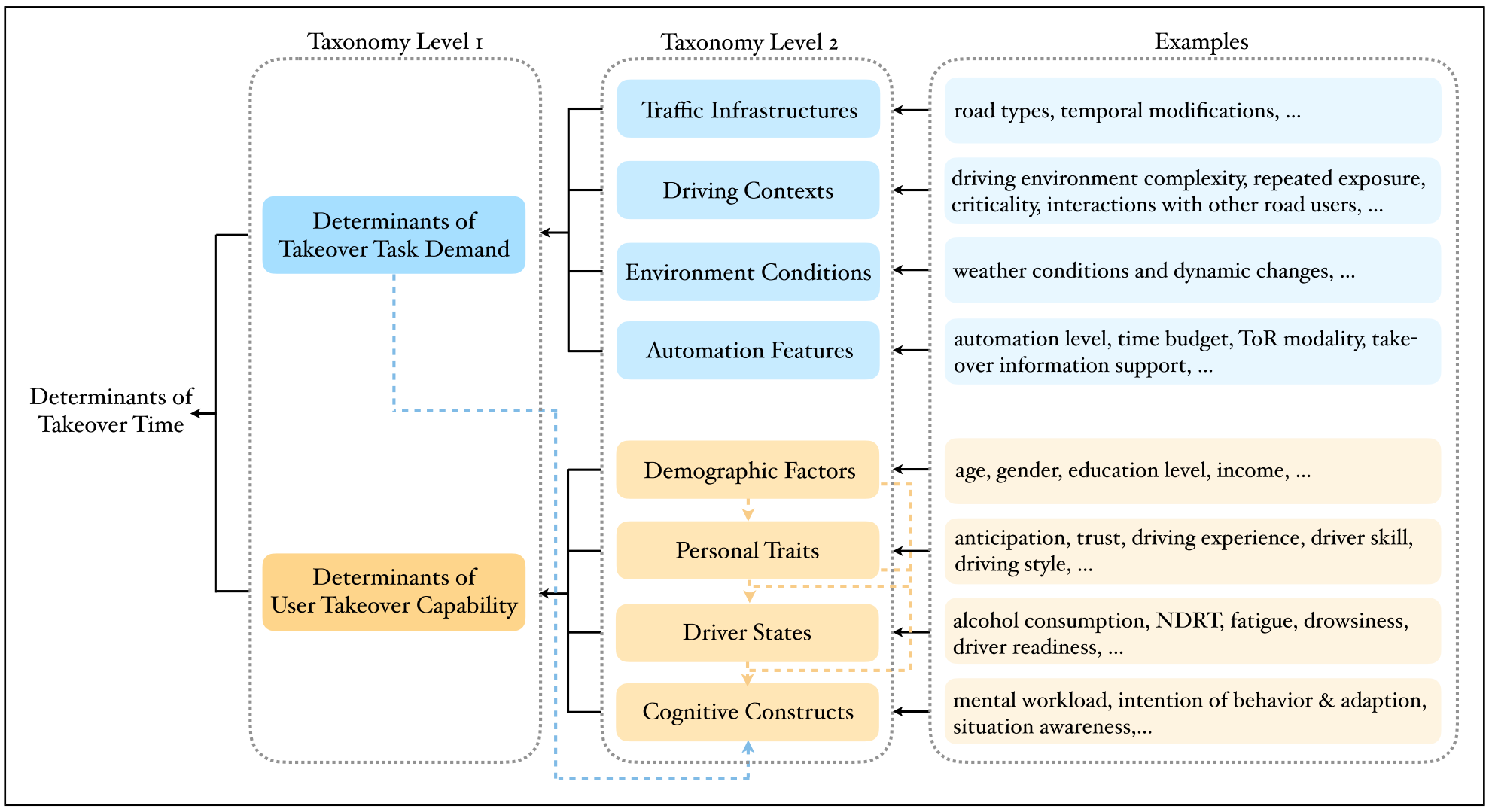

**Figure 4.** Taxonomy of takeover time determinants

## Taxonomy of Takeover Time Determinants

We propose a structured taxonomy to organize the identified determinants. This not only brings conceptual clarity but also lays the foundation for investigating their interrelationships and complex effects on takeover time. To ground our taxonomy, we adopt the Task-Capability Interface (TCI) model (Fuller, 2011), a well-established framework frequently used to explain driver behaviors (Calvert et al., 2020; Delmas et al., 2022; Oviedo-Trespalacios et al., 2019). The core concept is that drivers adjust their behavior to manage perceived risk based on their assessment of task demands and their own capabilities (Fuller, 2011; Van Lint & Calvert, 2018). In takeover contexts, we apply this model to categorize determinants of takeover time into two groups: takeover task demand determinants and driver takeover capability determinants, as shown in Figure 4.

The determinants of takeover task demand reflect the requirements for various components of takeover scenarios. We modify the six-layer model for automated driving scenarios in Weber et al. (2019) and make the classification model suitable for takeover scenarios. Thus, a four-level classification for determinants of takeover task demand is proposed, including traffic infrastructures, driving contexts, environment conditions, and automation features. The determinants of driver takeover capability reflect drivers' competences in fulfilling takeover tasks. Fuller (2011) divided driver capability into related knowledge, skill, and other human factors, but this classification is ambiguous in the takeover context. Hence, we refer to the framework of human factors in traffic modeling in Sharma et al. (2018) and reclassify the capability determinants into four categories: demographic factors, personal traits, driver states, and cognitive constructs. We anticipate that the proposed taxonomy of takeover time determinants can help to elucidate the underlying mechanisms that govern drivers' takeover behaviors and facilitate the prediction of takeover time which is the lower limit to be satisfied by sufficient time budgets.

## Taxonomy of Takeover Outcome Indicators

A variety of outcome measures has been used in studies of takeovers (Alrefaie et al., 2019;

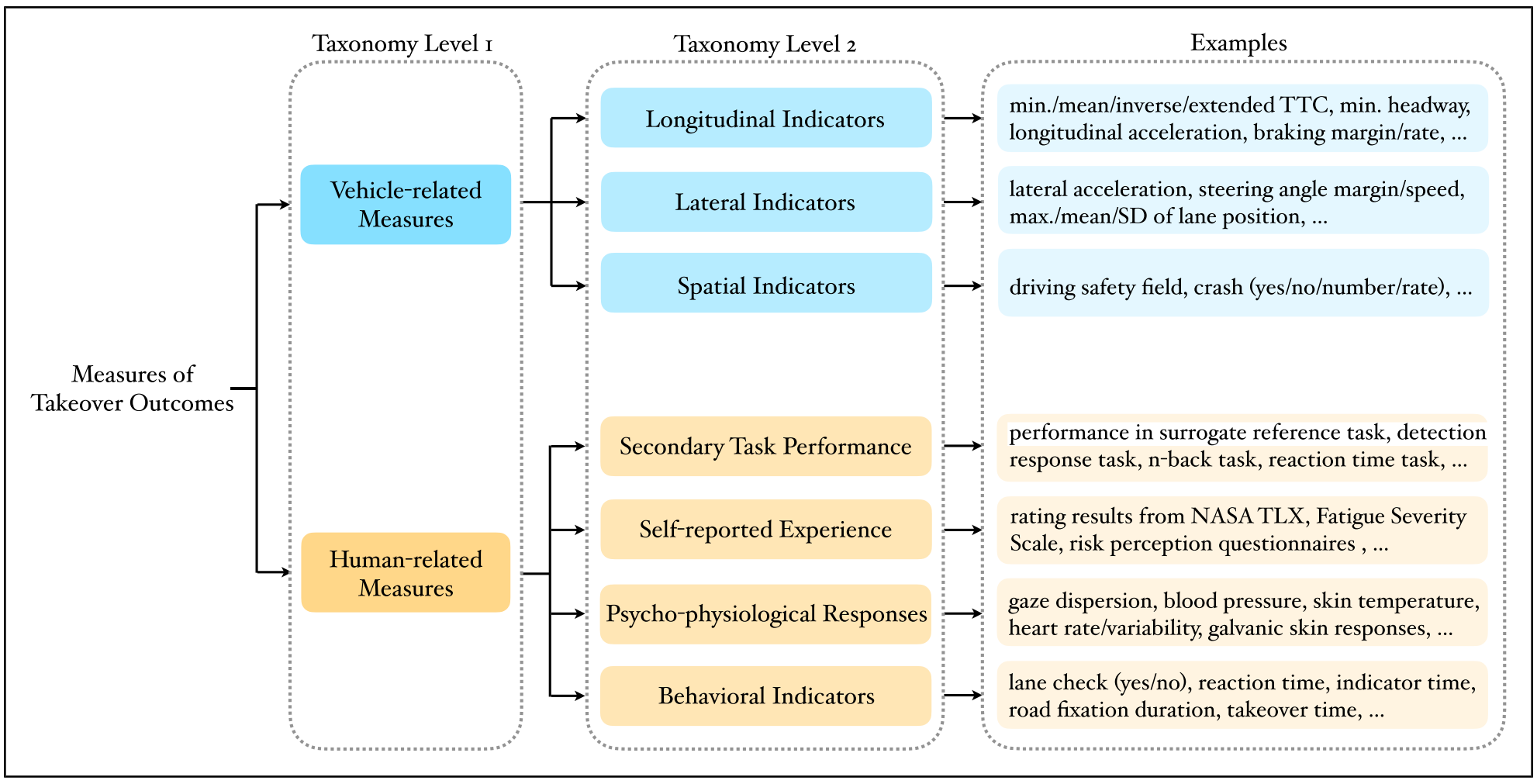

**Figure 5.** Taxonomy of takeover outcome measures

Eriksson & Stanton, 2017a). However, a standardized and agreed-upon criterion for measuring takeover outcomes is still missing (Li et al., 2023). This makes it difficult to consolidate the results across studies and draw consistent conclusions. It is necessary to provide an overview of the existing measures of takeover outcomes, which can help to clarify the suitability of the evaluation criteria and guide the development of new measures of takeover outcomes.

This review proposes a taxonomy to categorize measures of takeover outcomes as shown in Figure 5. Given that takeover is an interaction between vehicles and humans, we divide the measures of takeover outcomes into vehicle-related and human-related measures. Specifically, vehicle-related measures are based on driving parameters, such as lateral acceleration (McDonald et al., 2019). And human-related measures are based on individual behaviors and subjective feedback, such as heart rate variability (Hecht et al., 2018). These two categories are not contradictions but complement each other. Moreover, we propose to divide vehicle-related indicators into longitudinal, lateral, and spatial indicators according to their application scope, that is, the specific dimensions of takeover performance that these indicators measure. Here, spatial indicators refer to the metrics that comprehensively measure the takeover performance from both longitudinal and lateral dimensions, such as the driving safety field (Wang et al., 2015). Collision-related metrics are also categorized into spatial indicators. This is because collision is the result of operational failures in both longitudinal and lateral dimensions, that is, drivers fail to either brake or change lanes in time to avoid crashes. As for human-related indicators, we propose to further divide them into four categories according to the types of data that are collected and analyzed, namely, secondary task performance, self-reported experience, psycho-physiological responses, and behavioral indicators. Note that even though some behavioral indicators are measures based on vehicle-related parameters (such as steering wheel angles), human cognitive activities and the corresponding behaviors are the key determinants of these indicators. Therefore, we categorize such behavioral indicators as part of human-related measures rather than vehicle-related measures.

This taxonomy can provide a comprehensive understanding of the measures of takeover outcomes, guide the development of standardized outcome measures, and then improve the comparability of studies of takeovers. We suggest that a more balanced measure that integrates vehicles' objective driving quality and human drivers' subjective experience is essential for the development and widespread adoption of conditionally automated driving systems.

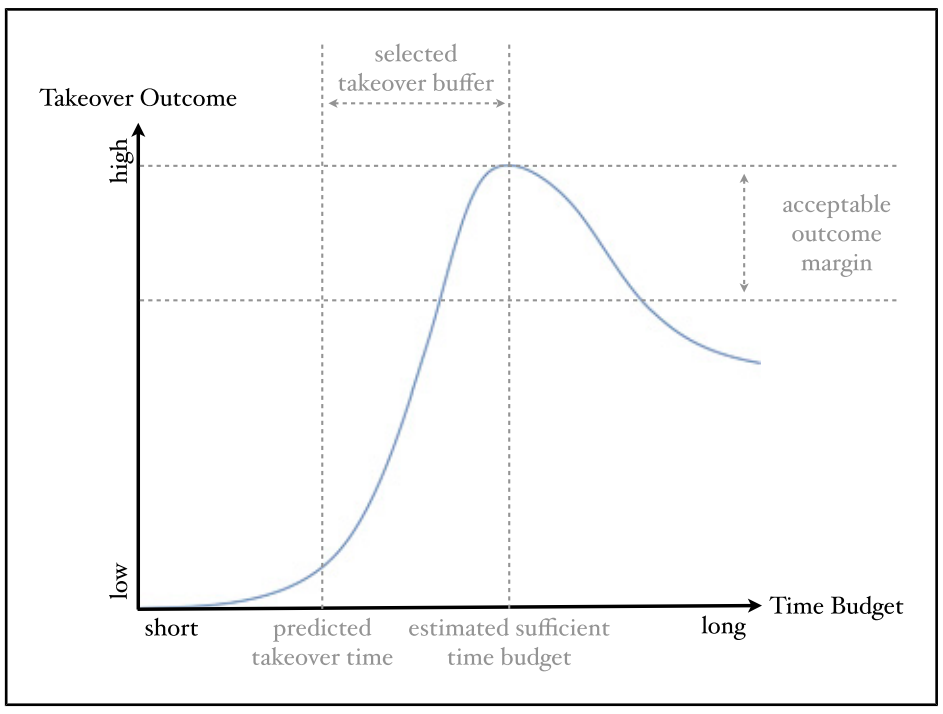

**Figure 6.** Qualitative relationship between takeover time, time budget, and takeover outcome

## *Qualitative Relationship Between Takeover Time, Time Budget, and Takeover Outcomes*

The sufficiency of the time budget is a relative concept of the offered time budget and the needed takeover time, which can be reflected by the corresponding takeover outcomes (Tanshi & Söffker, 2022). In this case, research on sufficient time budgets requires a clear quantitative relationship between takeover time, time budget, and takeover outcomes. It can guide conditionally automated driving systems to accordingly adjust the time budget for the predicted takeover time to optimize the takeover outcomes.

We propose a qualitative relationship between takeover time, time budget, and takeover outcomes in Figure 6. This serves as a starting point for future research on sufficient or adaptive time budgets. The vertical lines are illustrative examples of a predicted takeover time and an estimated sufficient time budget. The inverted U-shaped relationship shown in the figure pertains to this specific takeover time. Note that the takeover outcome here embodies not only objective assessments of the safety and quality of takeovers, but also drivers' subjective assessments of experiences and feelings. As shown in Figure 6, conditionally automated driving systems first predict a specific takeover time that human drivers need. If the offered time budget is much shorter than the takeover time, the takeover outcome is generally poor because drivers do not have enough time to resume control and return to their normal driving level (Gold et al., 2018). This situation can lead to over-reactions and even collisions in the worst case (Gold et al., 2013). Then, with the increase of the time budget, the takeover outcome will improve first because drivers have more time to respond and develop situation awareness (Endsley, 2021). Consequently, the safety, quality, and comfort of the takeover process will be enhanced (Wan & Wu, 2018; Yin & Pan, 2020) during which process the time budget gradually exceeds the takeover time. After the takeover outcome rises to a peak, the time budget will become unnecessarily long as it continues to increase. The safety and quality of the takeover will not be improved significantly due to the limitations of driver takeover capabilities (Huang & Pitts, 2022; Wan & Wu, 2018). On the contrary, the overall takeover outcome will gradually drop, as the takeover process becomes inefficient which can harm drivers' takeover experience (Gold et al., 2013; Zhang et al., 2019). Considering the safety and quality of takeover are maintained at a high level, the takeover outcome declines at a moderate rate after the peak.

While this roughly inverted U-shaped relationship has been acknowledged in several studies, Figure 6 provides new insights from the following three aspects:

**Highlights the role of the takeover buffer:** We emphasize the essential role of the takeover buffer in the interplay between takeover time, time budget, and takeover outcome. For a predicted takeover time within a specific context, selecting an appropriate takeover buffer is crucial to defining a time budget that is estimated to be sufficient to achieve the desired takeover performance. Therefore, the selection of a takeover buffer is fundamental to ensuring that the time budget adequately supports the takeover task.

**Proposes an acceptable outcome margin:** Rather than focusing solely on the peak point where outcome may decline with an extended time budget, we suggest that research explore establishing an acceptable margin for takeover outcome that a substantial portion of drivers can reliably achieve. This approach ensures that any reduction in takeover outcome due to a prolonged time budget remains within an acceptable threshold. Within this margin, time budgets can be adjusted to optimize user experience without compromising essential performance standards.

**Extends to a three-dimensional relationship model:** We argue that the relationship between time budget and takeover outcome may shift with varying takeover times. Based on suggested time budgets and takeover times (see Table 2), we find that drivers tend to prefer longer buffers as their takeover time increases. This may occur because longer takeover times often accompany more complex tasks, which demand additional buffers for a safe and effective takeover. Consequently, peak takeover outcome and the curve's shape may adjust with different takeover times. Zhang et al. (2019) supported this view, noting that drivers take longer to resume control when provided with more time. This two-dimensional model of time budget and takeover performance can be expanded to three dimensions by incorporating takeover time. Additional research is necessary to further test and quantify this model for safe and comfortable takeovers.

These considerations contribute to a more refined understanding of the relationship between takeover time, time budget, and takeover outcome, outlining how adaptive time budgets can be tailored to diverse driving scenarios. To empirically quantify the inverted U-shaped relationship of takeover time, time budget, and takeover outcome, this study proposes a structured research roadmap encompassing the following three avenues:

(1) Controlled experiments with varying time budgets:

This is the primary empirical methodology required to generate the data for validating the proposed hypothesis. Because the qualitative relationship includes extremely short time budgets that may pose safety risks, we recommend conducting experiments in high-fidelity driving simulators. The tested time budgets can be systematically manipulated from very limited to excessively long, for example, by selecting values proportional to drivers' takeover times reported in prior studies under comparable conditions. Additional experimental manipulations, such as varied driving scenarios or secondary tasks, may be used to elicit a range of takeover times. This would enable a richer set of combinations of takeover time and time budget for validation. Such designs, however, require careful attention to resource allocation and order effects.

(2) Composite takeover outcome assessment:

Validating and quantifying the qualitative relationship requires a composite outcome metric that integrates both objective measures (e.g., reaction time, minimum time-to-collision, and steering entropy) and subjective evaluations (e.g., workload, comfort, and satisfaction). The weighting of different metrics may be chosen based on design goals or participant preference judgments (which is similar to pairwise weighting procedures used in the NASA Task Load Index).

(3) Model identification:

With experimental data and composite outcome scores, the mathematical function that best characterizes the relationship among takeover time, time budget, and takeover outcome can be identified. Statistical methods can be applied, such as mixed-effects regression, generalized additive models, and Bayesian nonlinear modeling. Machine learning methods can also be used, such as neural additive models and gradient boosting machines. These predictive models are expected to quantify the inverted U-shaped pattern and can be used to select the optimal time budget associated with given takeover time.

The implementation of adaptive time budgets need to consider the legal and ethical consequences of time budget miscalibration. A time budget implicitly encodes assumptions about driver capability of fulfilling the takeover task and the corresponding task demand. When these assumptions are miscalibrated, particularly when systems provide insufficient time budgets, they can lead not only to technical performance degradation but also to ethical harms and legal exposure. Specifically,

(1) From an ethical perspective, miscalibrated time budgets raise questions about the duty of care owed by automated driving systems to provide drivers with proper opportunities to regain vehicle control (Bergmann, 2022). Insufficient time budgets may induce panic and reduce trust, thereby violating the principle of nonmaleficence by exposing drivers to preventable harm. Overly conservative time budgets can reduce engagement, distort expectations, and foster automation complacency, weakening the driver's role as a competent moral agent and fallback. To mitigate these risks, ethical deployment requires transparency (Rowe et al., 2024): time budgets should be communicated as uncertain estimates, and systems should clearly convey their limitations and associated risks to support informed driver action.
(2) From a legal perspective, miscalibrated time budgets complicate post-incident liability attribution (Diaz-Piedra et al., 2023), as unsafe outcomes may result from system design decisions rather than driver behavior alone. An insufficient time budget can reflect both a system-level failure to provide a safe exit strategy for the automation and a situation in which drivers are not fully prepared for, or fail to adequately monitor, the driving task. Conversely, if a system consistently provides excessive buffers and a crash occurs, a manufacturer may argue that the driver was negligent in responding to the takeover request, whereas the driver may contend that the system's over-calibration fostered automation complacency. Therefore, legal disputes will rely heavily on event data recorders (Martinesco et al., 2019). Standardizing the log interpretation is crucial.

These challenges highlight the need for research on transparent, auditable, and evidence-based time budget design. Integrating ethical and legal considerations into future studies can help ensure that adaptive time budget models support not only technical performance but also accountability, trust, and socially responsible deployment.

## Conclusions

Despite the acknowledgment of time budgets as a crucial factor, research specifically focused on determining what constitutes a sufficient time budget remains limited. Studies have concentrated on minimizing takeover time, often overlooking the complexities of achieving a balance between timely control transitions and effective performance. This gap in the literature prompts a fundamental question: How can sufficient time budgets be established for safe and comfortable vehicle control transitions across various drivers and scenarios? To answer this open question, this study considers the takeover process as an integrated sequence: takeover time serves as the lower limit of sufficient time budgets and takeover outcomes are the consequences of the supplied time budgets.

We systematically examine the research on the takeover time, time budget, and takeover outcomes in conditionally automated driving. Results show that drivers' takeover time varies significantly due to its complex causal relationships with numerous determinants. The fixed time budget is sub-optimal in meeting various takeover demands compared with the adaptive time budget. We propose a qualitative relationship between takeover time, time budgets, and takeover outcomes. The proposed taxonomies of takeover time determinants and takeover outcome measures provide a promising research avenue for research on adaptive time budgets. Rather than prescribing fixed time budget values, we establish a flexible and adaptable framework that estimates sufficient time budgets by combining the predicted takeover time with an appropriate takeover buffer to achieve desired outcomes. Our contribution lies not in a precise calculation but in providing a structured method adaptable to a variety of contexts. This offers a foundation for future research on adaptive time budgets for safe and comfortable takeovers.

## Key Points

- This is a systematic and extensive review synthesizing the literature on takeover sequences, discussing key aspects such as takeover time, time budgets, and takeover outcomes.
- A taxonomy of takeover time determinants is proposed to guide the selection of appropriate predictors.
- Another taxonomy of takeover outcome indicators is proposed to offer insights for developing standardized frameworks of performance measures.
- A qualitative relationship between takeover time, time budget, and takeover outcome is made as a preliminary foundation for research on designing sufficient time budgets.

## ORCID iD

Kexin Liang https://orcid.org/0000-0002-8417-3974

## Funding

The authors received no financial support for the research, authorship, and/or publication of this article.

## Declaration of Conflicting Interests

The authors declared no potential conflicts of interest with respect to the research, authorship, and/or publication of this article.

## Author Biographies

**Kexin Liang** is a PhD candidate in the Department of Transport & Planning at Delft University of Technology (TU Delft), Delft, the Netherlands. She received her master in control science and engineering from Beijing Jiaotong University in 2021.

**Simeon C. Calvert** is an associate professor of Smart & Automated Driving in the department of Transport & Planning at the TU Delft. He is the director and founder of the Automated Driving & Simulation (ADaS) research lab and co-director of the CityAI-lab for research on urban behavior using AI. He is also a board member of the Centre for Meaningful Human Control over AI, the interdisciplinary research program for responsible autonomous technology. He received his PhD in probabilistic macroscopic traffic flow modeling from TU Delft in 2016.

**J. W. C. van Lint** is a full professor of Traffic Simulation and Computing in the Department of Transport & Planning at the TU Delft. He received the M.Sc. degree in civil engineering informatics and the Ph.D. degree in transportation from Delft University of Technology (TU Delft) in 1997 and 2004, respectively. He was an Information Analyst and a Transport Engineer at various organizations. After receiving the PhD degree, he was appointed as an Anthonie van Leeuwenhoek (AvL) Full Professor by the Executive Board of TU Delft in 2013 (an honor reserved for only a few young scientists and educators). He is currently the Director of the Digitisation & AI For Mobility Network Dynamics Laboratory (DAIMoND Lab) in the Faculty of Civil Engineering and Geosciences. He is active in many international projects and collaborations.